\documentclass{ws-p8-50x6-00}

\begin{document}

\title{Summary Report of the\\[1ex] Spin Physics Working 
Group \footnote{Summary talk presented at the 9th
International Workshop on Deep Inelastic Scattering (DIS 2001), 
Bologna, April 2001.}}

\author{Nicola Bianchi}

\address{INFN - Laboratori Nazionali di Frascati,\\
via E. Fermi 40, I-00044 Frascati, Italy\\ 
E-mail: bianchi@hermes.desy.de}

\author{Rainer Jakob}

\address{Fachbereich Physik, Universit\"at Wuppertal,\\ 
42097 Wuppertal, Germany\\
E-mail: rainer@theorie.physik.uni-wuppertal.de}  

\maketitle

\abstracts{
The contributions to the Spin Physics WG are summarized. Several new
experimental results and plans for new measurements have been reported.
An improved theoretical understanding of the most recent hot topics
in spin physics  has been discussed by many authors especially
in the new fields of the transversity and generalized parton distributions.}

\section{{\large DIS$_{2001}$}-claimer}
The spin physics working group at the DIS 2001 workshop hosted a total 
of 37 talks equally distributed between experimental and theoretical 
reports. The topics covered many different aspects of polarization phenomena 
in deep inelastic scattering. This summary presents
recent developments reported at the workshop, and sketches the contours 
of the ongoing world wide spin program in hadron physics. Unavoidably, a 
summary cannot reflect all facets and details of such a variety of
contributions in a fair manner. Sorry~!  

\section{Introduction}

Spin -- merely an inessential complication in particle physics ? No, anything
but that ! On the contrary, over and over again spin physics proves to be a
vital element and indispensable tool in our venture of revealing the internal
structure of hadronic matter. Once again, at this workshop, the numerous
contributions, above average attendance, and lively discussions in the
parallel sessions of the spin physics working group testified the vividness of
the field. The physics topics were divided in five different 
subsections, namely:  

\begin{itemize}
\item Spin structure functions $g_1$ and $g_2$
\item Polarized parton distributions
\item Single-spin azimuthal asymmetry and transversity distribution
\item Generalized (skewed) parton distributions and exclusive processes
\item Starting experiments and new projects
\end{itemize}

From the above list of topics, it is quite  evident that spin physics 
is in a rich, exciting and dynamical situation and  that a new era, 
which investigates new types of fundamental distribution functions, is
started, while the first era is still far from being exhausted.

After the early measurements performed at SLAC, the
modern era in spin physics begun in 1988 when the EMC collaboration
reported the experimental evidence that only a small fraction
of the proton spin is carried by the quarks\cite{Ashman:1988hv}.
This shocking result has produced an enormous experimental effort during the
90s when several measurements have been successfully
performed at CERN, SLAC and DESY with a precise determination of
the spin structure functions. To better underline the 
importance of the EMC measurement is sufficient to
mention that the related publication,
together with the old ones on the $J/\psi$ discovery\cite{Aubert:1974js} and 
with the recent one on
the atmospheric neutrino oscillation\cite{Fukuda:1998mi}, is the most cited
experimental paper in high energy and nuclear physics
of the last 30 years\cite{SPI}.

From inclusive and semi-inclusive polarized data, the helicity
distribution for the {\em up}- and {\em down}-quarks are now known
with reasonably good precision, while still very sparse informations
are  available for the sea-quark and gluon helicity
distributions. These quantities are planned to be measured more
precisely in several experiments at DESY, CERN, RHIC and 
SLAC\cite{Filippone:2001ux}.

Although the importance of the leading twist transversity distribution 
$h_1(x)$ (often also denoted $\delta q(x)$) for completing the knowledge on
the spin structure of the nucleon was recognized quite some time 
ago\cite{Ralston:1979ys}, only recently the experimental hunt for it 
started. Unlike the other two leading-twist
distribution functions $f_1(x)$ and $g_1(x)$ (or $q(x)$ and $\Delta q(x)$), 
the transversity distribution function is still unknown due to its 
chiral-odd nature\cite{Barone:2001sp}. First HERMES and 
SMC data\cite{Airapetian:2000tv} showed that
a non zero spin dependent T-odd fragmentation function (the Collins
f.f.) allows to access the transversity distribution and even higher-twist
helicity changing distributions. More options to unravel transversity 
-- each with its own advantages and caveats due to the subtle interplay of 
chirality and time-reversal issues -- are under current investigation.

Finally the recently developed formalism of Generalized Parton 
Distributions (GPDs) for a description of the exclusive 
leptoproduction reactions  has deepened the insight in the connection between
exclusive and inclusive hard reactions. Moreover, accessing GPDs 
experimentally appears to be a promising
way to acquire additional and complementary informations on the spin 
structure of the nucleon \cite{Goeke:2001tz}.
Due to the experimental difficulties to measure exclusive reactions, 
like Deeply Virtual Compton Scattering (DVCS),
at sufficiently hard scale, first results in this field are appearing
just now\cite{Airapetian:2001yk}.
 
\section{Spin structure functions $g_1$ and $g_2$}

The spin structure function $g_1$ is now well determined for
proton, deuteron and neutron. Most of the available data cover the 
DIS region with $Q^2 > 1$ GeV$^2$ and $W > 2$ GeV.
Unlike the unpolarized structure functions which have been measured
over a broad kinematic range, no polarized data are available
for large $Q^2 > 70$ GeV$^2$ or small $x < 10^{-3}$. 

\begin{figure}[ht]
\epsfxsize=20pc 
\hspace{1.0cm}\epsfbox{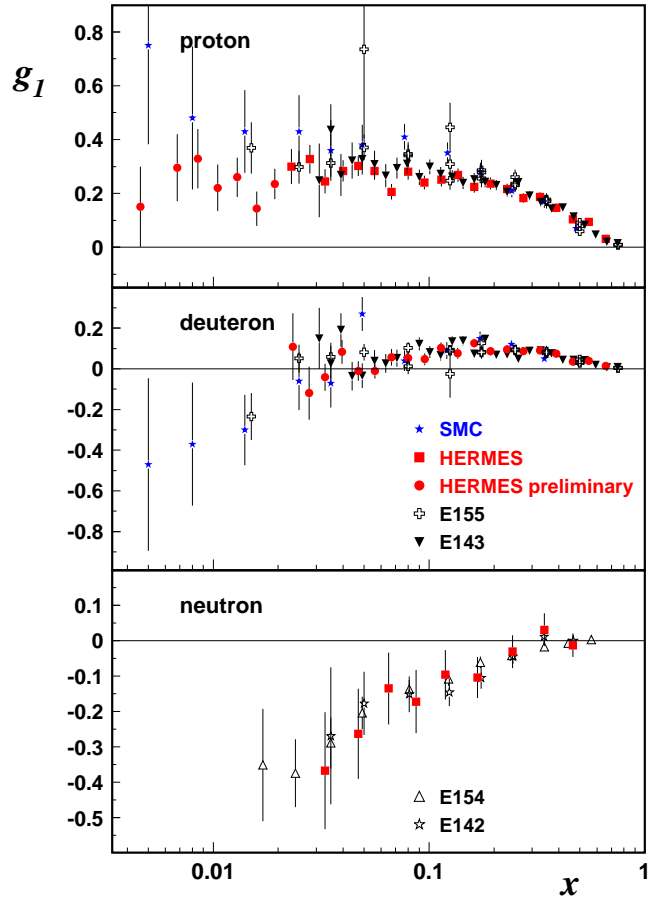} 
\caption{World data on spin structure function $g_1$ collected during the 90s. 
All the data are shown at the measured $Q^2$. 
SMC results for $Q^2 < 1$ GeV$^2$ at very low $x$ are not shown.  
\label{fig:g1p}}
\end{figure}

In fig.~\ref{fig:g1p} the world data for $g_1$ in the DIS region
are presented. Together with the already published data, the preliminary
HERMES data 
(see contribution of {\it C. Weiskopf}) for $g_1^p$ at low-$x$ and low-$Q^2$
and for $g_1^d$ are also shown. It is worth to note that the latter data set
represents only a small fraction ($\sim$1/6) of the already collected data on
the deuteron. 
An important experimental finding is the $Q^2$-independence of the inclusive
asymmetry $A_1 \sim g_1/F_1$ in the whole measured kinematic range. This
suggests a similar behavior of the scaling violation for the structure
functions $F_1$ and $g_1$.
After the completion of the analysis of all HERMES data on the deuteron,
no substantial improvement of the knowledge of $g_1$ in the DIS region of
moderate $x$ and $Q^2$ is foreseen in the near future.

On the contrary, several measurements are ongoing  or are planned in the
photoproduction or in the resonance electroproduction 
regions (see {\it S. Kuhn}). 

\begin{figure}[t]
\epsfxsize=20pc
\hspace{2.0cm}\epsfbox{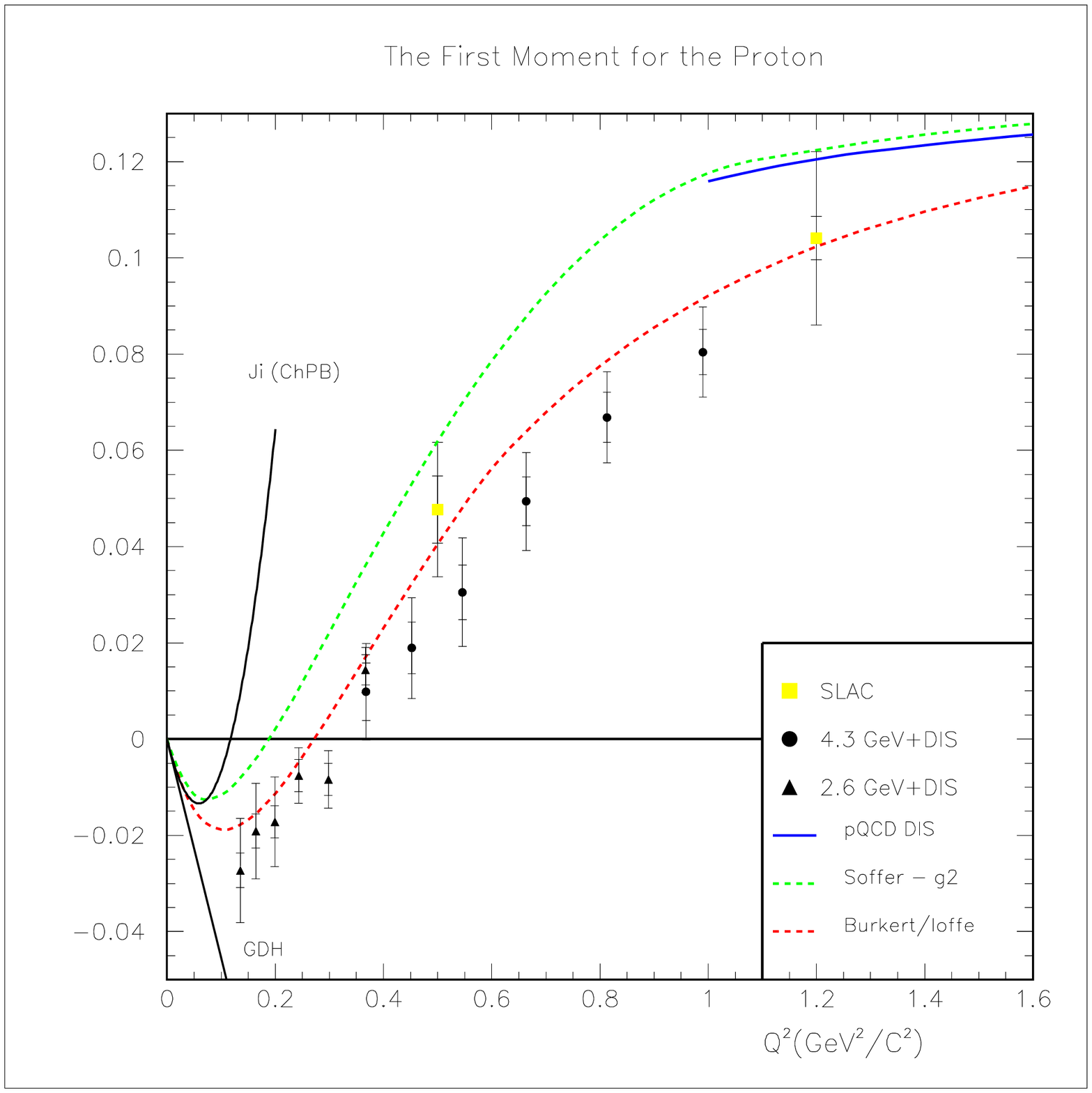} 
\caption{Preliminary results for the first moment
of $g_1$ for the proton as a function of $Q^2$ from CLAS
at Jefferson Lab. Also shown are the data from SLAC E143 and different
predictions.
\label{fig:int}}
\end{figure}

Unlike the unpolarized case, there are no results for the spin-dependent
cross section for real photon for which only phenomenological
extrapolations from the DIS  region are 
available\cite{Bianchi:1999qs}. First preliminary
data collected at MAMI and ELSA have been reported (see {\it K. Helbing})
which show the dominance of the helicity 3/2 in the polarized photoabsorption,
due to the $P_{33}$, $D_{13}$ and $F_{15}$ resonance excitation.
The GDH collaboration is planning to measure the spin dependent cross sections
for the proton and the neutron up to 3 GeV in order 
to explore the onset of the Regge regime.
In addition a new polarized real photon beam facility has been
approved at SLAC (see {\it G. Peterson}) which will extend this measurement
up to 40 GeV.

In the nucleon resonance region preliminary results are available from
JLab. In fig.~\ref{fig:int} the first moment $\Gamma_1$ of the spin
structure function $g_1$ for the proton is shown 
as measured by the CLAS Collaboration (see {\it G. Dodge}). 
The data show a strong $Q^2$-dependence of the integral which becomes
negative at very low $Q^2$ due to the role of the $P_{33}$ ($\Delta$-resonance)
in this domain. Also reported (see {\it W. Korsch}) is the 
Hall-A Collaboration measurement of generalized GDH integral of the neutron as 
function of $Q^2$ which approach the Sum Rule expectation value for the real
photon. 
Due to the limited energy of the JLab beam, both $\Gamma_1$ and GDH integrals
at low $Q^2$ include a high energy extrapolation which determination will
strongly benefit from the planned real photon measurements.

\begin{figure}[t]
\hspace{1.0cm}\includegraphics[clip,width=8cm,height=8cm]{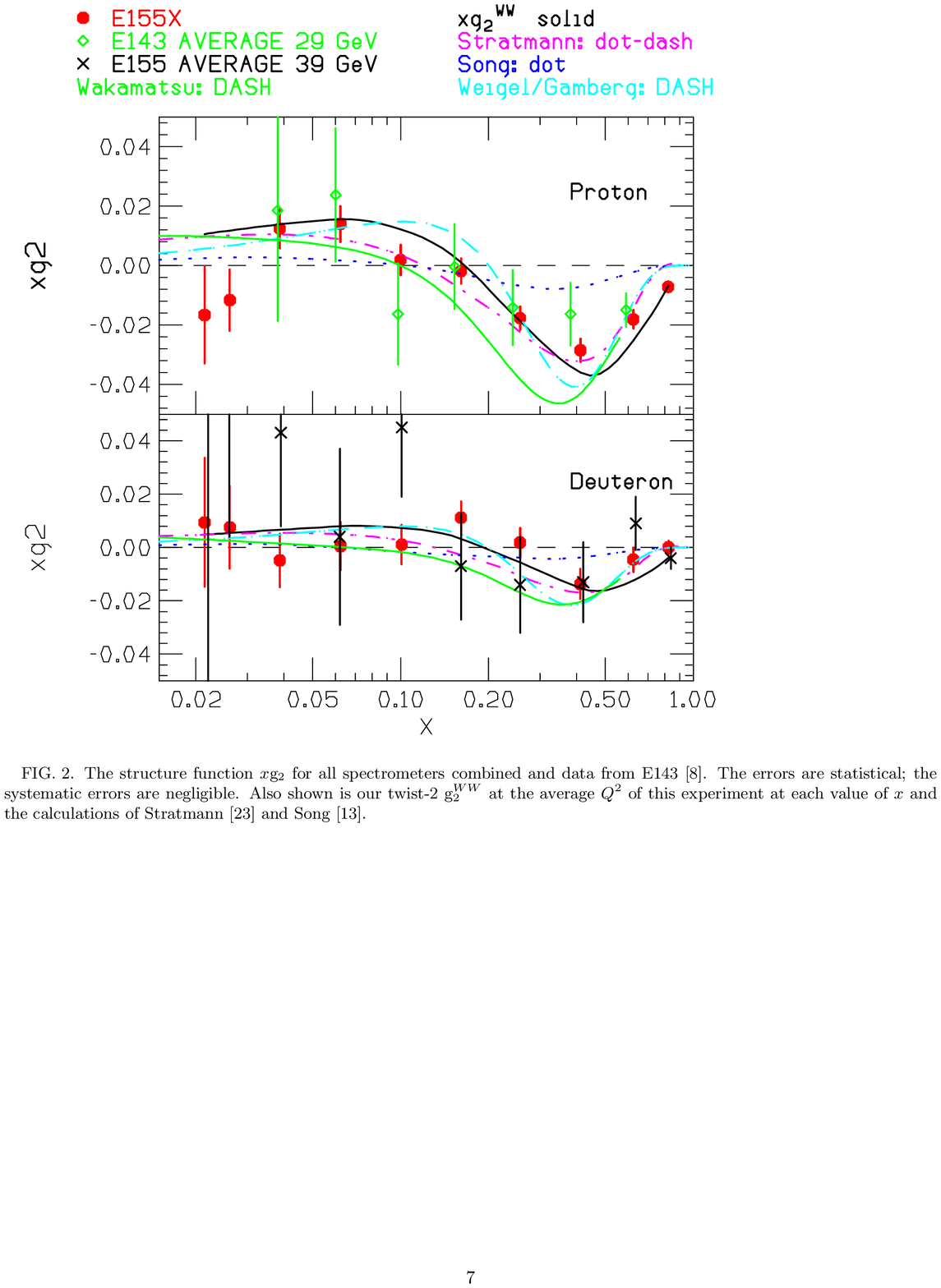}
\caption{SLAC data on spin structure function $g_2$. Close circles are the
  preliminary E155x results. Curves are predictions from Stratmann, Weigel,
  Wakamatsu and Song.  
\label{fig:g2}}
\end{figure}

New and precise results on the spin structure function $g_2$ from 
the SLAC E155x collaboration (see {\it O. Rondon}) show for the first 
time a possible
small deviation from the dominant twist-2 contribution (Wandzura-Wilczek
term) for the proton. The available SLAC data\cite{Abe:1996dc} on
$g_2$ are shown in fig.~\ref{fig:g2}. These data allow to extract the
twist-3 reduced matrix element $d_2$ which appears to be slightly positive for
both proton and neutron. This result is in contradiction with QCD sum rule
models while is in better agreement with quark bag and chiral soliton models.
Like for the structure function $g_1$ case, no further experimental
improvement is foreseen for $g_2$ in the DIS regime in the near future. 

\section{Polarized parton distributions}

\begin{figure}[t]
\epsfxsize=20pc
\hspace{1.0cm}\epsfbox{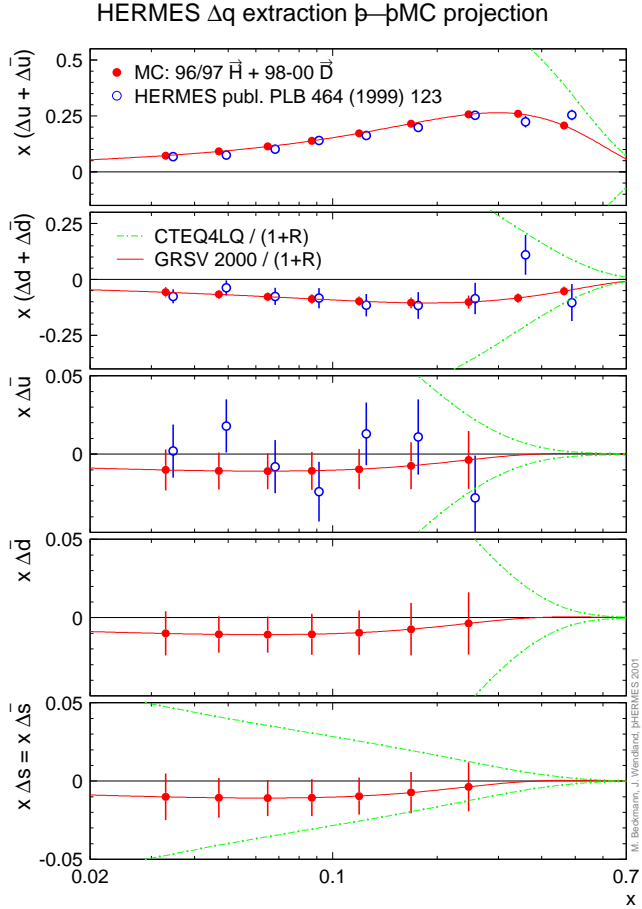} 
\caption{Expected precision for polarized quark distributions from the 
inclusive and semi-inclusive data
already collected by HERMES and with no flavor symmetry assumption on the sea
quarks. Also shown are the published HERMES results based on 1995-1997 runs.
\label{fig:deltaq}}
\end{figure}

Inclusive results for spin structure functions allow to extract 
informations on the quark and
the gluon helicity distributions through NLO QCD fits.
A 8 parameter fit of $A_1$ data, performed in the $\overline{MS}$ scheme
and under the SU(3) assumption, has been presented (see {\it J. Bl\"{u}mlein}).
In this fit, the parameter correlations were fully taken into account in the
Gaussian error propagation. Another QCD fit of $A_1\sim g_1/F_1$ data in the
$JET$ scheme has been showed\cite{Leader:1998qv} 
(see {\it D. Stamenov}). The advantage of fitting
$A_1$ instead of $g_1$ results in the cancellation of higher twist effects for
$Q^2$ down to 1 GeV$^2$.
These new fits confirmed that inclusive data alone poorly constrain the
separate quark and anti-quark polarized distributions and the gluon
distribution as well (see {\it A. Miller}).

Combined inclusive and semi-inclusive QCD analysis have been
performed\cite{deFlorian:2000bm} and showed that semi-inclusive 
data\cite{Adeva:1998qz} are in
perfect agreement with the inclusive data with respect to the determination of
polarized parton densities, but constitute a very powerful tool to study the
sea quark polarization ($\Delta\overline{u}$, $\Delta\overline{d}$ 
and $\Delta s$).
An updated analysis of the HERMES results with the inclusion of the
preliminary deuteron data has been shown (see {\it M. Beckmann}). This
analysis, based on the purity formalism, is still based on a symmetric
assumption of the sea.  
Since the unpolarized parton densities of the sea quarks are not flavor
symmetric, there is no reason to believe that this symmetry should hold for
the polarized case. Indeed there is a suggestion based on a meson-cloud model
that $\Delta\overline{u} -\Delta\overline{d} < 0$ (see {\it M. Miyama}). 
The full analysis of HERMES data with the inclusion of the large statistics on
the deuteron and the particle identification with the RICH detector will allow
not only to strongly improve the statistical precision on the polarization
of the {\it up}- and {\it down}-flavors but in addition to determine the
polarization for all different sea flavors without symmetry assumption
(see fig.~\ref{fig:deltaq}). The polarized quark distributions will be
also studied in the near future at COMPASS with semi-inclusive hadron
leptoproduction and at RHIC through the parity violating processes
$u\overline{d} \mapsto W^+ \mapsto l^+ + \nu$ and 
$d\overline{u} \mapsto W^- \mapsto l^- + \nu$.   

\begin{figure}[t]
\epsfxsize=20pc
\hspace{1.0cm}\epsfbox{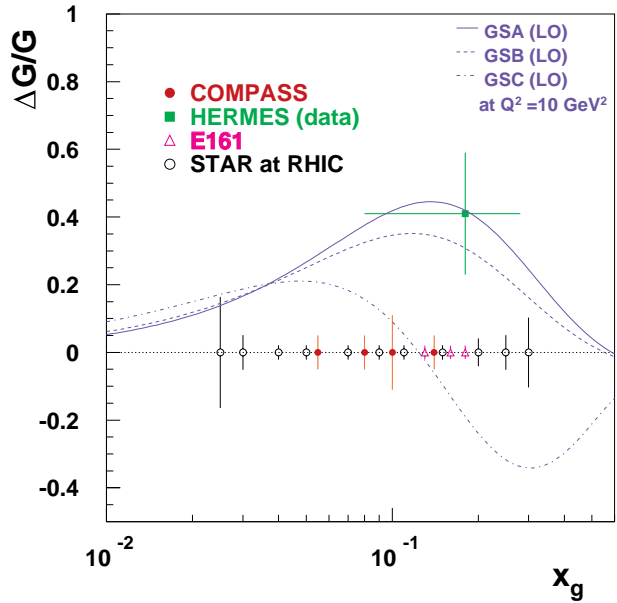} 
\caption{Expectations for future measurements of the gluon polarization. The
  HERMES result for high-$p_T$ hadron pairs is also shown. The curves are QCD
  analysis of polarized inclusive structure function.  
\label{fig:deltag}}
\end{figure}

One of the most important issues in spin physics for the next 5 years will be
the measurement of the gluon polarization which is poorly determined by
NLO QCD fits. 
Gluon polarization will be studied at COMPASS, SLAC and RHIC by measuring the
open charm production and the high-$p_T$ hadron pairs in the photon-gluon
fusion in $\vec{\gamma}+\vec{p}$ experiments and in the direct $\gamma$ 
production and two-jet events in $\vec{p}+\vec{p}$ experiments.
In addition running RHIC at very high energy ($\sqrt{s} = 500$ GeV)
will allow to access the gluon polarization through the
heavy quarks ({\em charm} and {\em bottom}) hadroproduction 
(see {\it I. Bojak}). 
In fig.~\ref{fig:deltag} the projected errors of the planned measurements 
are shown together with the first HERMES data\cite{Airapetian:2000ib} 
which interpretation
is improving by the better knowledge of all the competing sub-processes
(see {\it E.C. Aschenauer}).

COMPASS will start data taking this year using the previous SMC polarized
target (see {\it F.H. Heinsius}). Also the RHIC spin physics program will
start gradually this year and the full luminosity and the complete of the EM
calorimeter of the STAR detector will be reached in 2004 (see  {\it
  L. Bland}). SLAC experiment E161 is planned to run in 2003-2004 
(see {\it G. Peterson}).

There is a number of proposals for the medium-range future, where high 
luminosity experiments will strive for unprecedented precision in the
determination of polarized distribution functions and the gluon polarization
as some of their main goals: the TESLA-N experiment, or experiments at the
ELFE accelerator (see {\it D.~Ryckbosch}), as well as experiments at the
proposed EIC accelerator (see {\it A.~Desphande}) into which the former eRHIC
and EPIC proposals have merged.

\section{Single spin azimuthal asymmetries:\protect{\newline} 
Towards a global analysis of transversity}

Only the knowledge of the transversity distribution $h_1(x)$ will complete the
mapping of the spin structure of the nucleon at leading twist. Due to its
chiral-odd nature transversity is inaccessible in totally inclusive DIS; in 
order to form a chiral-even observable it
requires the presence of a second chiral-odd projection of a soft matrix
element in a more complex reaction -- a second transversity 
distribution, a higher twist distribution, or
an appropriate fragmentation function\cite{Jaffe:1997yz}. The object of 
desire will consequently appear in an observable
together with a second yet unknown function, usually in form of a convolution. 
Therefore, complementary experiments have to be utilized in a combined
analysis in order to finally extract the transversity distribution. 

This situation not only represents a considerable high level of
sophistication on the theoretical side, but also causes tremendous
demands on experimental setups. Thus, only now we are in the situation to 
seriously tackle unraveling of transversity. 

At this workshop to our knowledge truly the complete set of presently 
known ways to access transversity has been discussed in one or the 
other talk. The most promising approaches are sketched below.

\subsection{Collins effect in leptoproduction; a Single Spin Asymmetry}

An azimuthal asymmetry induced by the so-called Collins 
effect\cite{Collins:1993kk} is probably the most prominent representative 
of the different observables involving the transversity distribution. Its 
variant in 1-hadron inclusive 
lepton-nucleon scattering predicts a characteristic angular dependence 
in the azimuthal distribution of detected pions in the final state. The 
chain of logic is as follows: The transversity distribution 
describes how much of a transverse target polarization is transferred 
to a transversely polarized quark; the Collins fragmentation function, 
dubbed $H_{1}^\perp(z,k_T)$ in the 
{\it Amsterdam naming scheme}\cite{Mulders:1996dh}, describes 
the transfer of transverse quark 
polarization into preferred 
directions in the azimuthal distribution of produced hadrons, say pions.

\begin{figure}[ht]
\begin{center}
\includegraphics[width=0.96\textwidth]{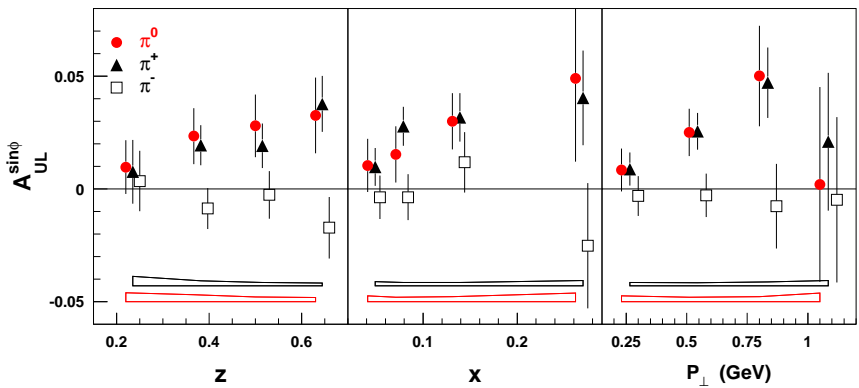}
\end{center} 
\caption{Single spin azimuthal asymmetry in electroproduction of pions in 
deep inelastic scattering (Collins effect) measured at HERMES. Shown is 
the $\sin(\phi)$ moment $A^{\sin}_{UL}$ for unpolarized beam and 
longitudinally polarized target vs.~$z$,$x$, and $p_T$ 
(see {\it D.~Hasch}). 
\label{fig:collins}}
\end{figure}

The observation of this asymmetry has been reported by HERMES and 
SMC\cite{Airapetian:2000tv}. In fig.\ \ref{fig:collins} the HERMES 
data are shown. Though
the target polarization longitudinal to the lepton beam on the average
corresponds to only small components of polarization transverse to 
the photon direction, the $\sin(\phi)$ dependence is rather 
convincingly confirmed for positively charged
and neutral pions (see contribution of {\it D.~Hasch}). A comparison with
theoretical estimates and limits shows good agreement 
(see {\it K.~Oganessyan}).

The effect of higher order perturbative corrections have to be taken into
account in an analysis of the asymmetries. For observables, which involve 
two scales like the Collins effect, there is a non-collinear factorization 
and the cancellation of double-logarithms is
incomplete. The remaining soft gluon corrections sum up to Sudakov suppression
factors. These Sudakov factors have been estimated to have significant 
effects on the magnitude of observed asymmetries\cite{Boer:2001he} 
(see {\it D.~Boer}). There is
also very recent progress on the understanding of the evolution of the
Collins fragmentation function, a question which is complicated by the fact 
that dependence on transverse momentum is to be kept. An analysis on this
issue, and the resulting evolution equation for $H_1^\perp$ in the 
large-$N_C$ limit was presented\cite{Henneman:2001ev} 
(see {\it P.~Mulders}). Both reported 
results constitute important progress towards a complete treatment of 
azimuthal asymmetries beyond the leading order.

\subsection{Transversely polarized $\Lambda$'s in a Double Spin Asymmetry}

Another opportunity to access transversity is to measure the transverse
polarization of produced $\Lambda$'s in semi-inclusive DIS. The
$\Lambda$'s are self-analyzing, i.e.~they reveal their spin by the subsequent 
decay into nucleon and pion, whose momenta have to be determined. 

The use of this possibility, which will become important with the improved
particle identification of future and starting experiments, relies on the 
knowledge of competing effects which also may result in transverse 
polarization of $\Lambda$'s. For instance, also unpolarized quarks can 
induce a transverse polarization of $\Lambda$'s depending on their 
azimuthal distribution in the jet\cite{Anselmino:2001vs}. This effect 
involves the only other T-odd leading-twist fragmentation 
function $D_{1T}^\perp(z,k_T)$, quasi the `mirror image' of the 
Collins f.f.~(see {\it F.~Murgia}).

Also a better understanding of the longitudinal spin transfer from quarks to 
$\Lambda$'s may shed light on this difficult issue (see {\it O.~Grebenyuk}
for experimental and {\it U.~D'Alesio} for theoretical aspects).

\subsection{Double Spin Asymmetries in semi-inclusive $\rho$ production}

Produced $\rho$'s in semi-inclusive DIS also reveal their spin via the
distributions of the decay products, i.e.~pion pairs. It turns out that
in principle there are three azimuthal asymmetries involving the transversity
distribution together with different spin-1 fragmentation 
functions\cite{Bacchetta:2000jk}
(see {\it A.~Bacchetta}). The involved fragmentation functions
can be extracted from the mass spectrum of pion couples; thus, they have to be
in close relationship with the two-pion f.f.\ discussed in the 
following paragraph.

\subsection{Single Spin Asymmetry in $\pi\pi$ production;\newline
interference fragmentation functions}

A very promising option to access transversity is given by a Single Spin
Asymmetry in the production of pion pairs in the same current 
jet\cite{Collins:1994kq}. The
relative momentum within the pair provides an additional lever arm to
extract the desired information. This allows for the integration over 
transverse momentum of the pion pair relative to the jet axis, which 
leads to a collinear factorization and the
vanishing of Sudakov suppression factors (see {\it D.~Boer}). 
The necessary T-odd and chiral-odd fragmentation functions are induced by
interference effects between two different channels for the production of 
the pion couple in the hadronization of the jet. The resulting 
azimuthal asymmetry in two-pion inclusive DIS was discussed, and a model 
estimate for the involved interference fragmentation function 
reported (see {\it M.~Radici}).

\subsection{Transversity perspectives}

The aforementioned options are currently under active theoretical 
investigation. First experimental observations of transversity related 
azimuthal asymmetries are reported, and a good share of future experimental 
efforts, for instance at HERMES and COMPASS experiments, and part of 
the RHIC spin program, will go into the `transversity project'. 
Experimental feasibility studies and projections appear to be very 
promising for the near future. 

For the extraction of the transversity distribution the analyses of
the asymmetries will have to be supplemented by independent knowledge on some 
fragmentation functions, for which $e^+e^-$ data
(for instance from LEP experiments, or data from the future $B$-factories) 
seem to be the best possible source of information.

Most likely, there will not be one {\it golden way} to the
extraction of the transversity distribution, but the need for a
{\it `global transversity analysis'} ({\it D.~Boer}). In fact, the 
enterprise has already started.

\subsection{Remark}
Sometimes the, well justified, excitement about the transversity
distribution as the missing spin distribution function obscures the 
clearness of the general aim: to collect as much as possible information 
and insight into the mechanism of confinement, and thus into the inner 
structure of hadrons. The knowledge on complicated spin dependent 
fragmentation functions, and for instance higher-twist distribution 
functions (see e.g.\ {\it Y.~Koike} for a theoretical approach to 
twist-three functions from gluonic poles\cite{Kanazawa:2001cx}), is not 
merely a prerequisite for a transversity analysis . The 
effort on extracting those functions is justified in its own interest 
as steps towards a full understanding of confinement at work.

\section{Generalized parton distribution and exclusive processes}

\subsection{Why should we bother to measure GPDs ?}

In fact, this is not the right question
to be asked, since it is practically impossible to avoid dealing with GPDs 
when utilizing hard reactions to unravel the inner structure of 
hadrons. Every form factor is an $x$-moment of a GPD; every ordinary
distribution function is given by a GPD in a certain kinematical 
limit. From this point of view GPDs have been studied at least for 
50 years. A more sensible question would be: What new aspects can we learn 
from GPDs, inaccessible with ordinary parton distributions and form 
factors~? 

\begin{figure}[htb]
\begin{center} 
\includegraphics[width=0.8\textwidth]{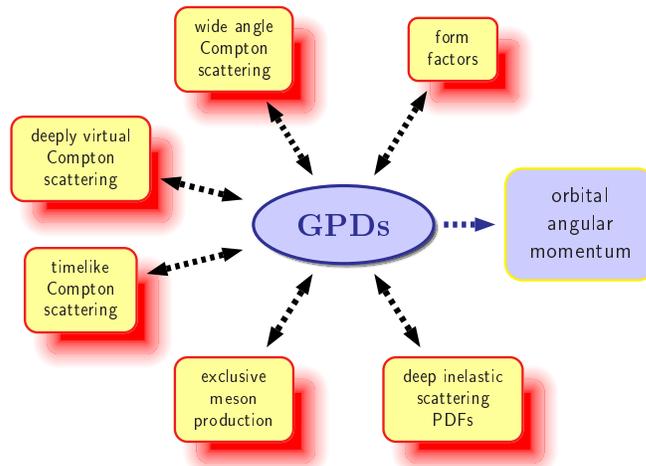}
\end{center} 
\caption{Generalized parton distributions occur in the description
of many different hard reactions; sometimes in the form of $x$-moments or in
special kinematical limits.\label{fig:gpd-map}}
\end{figure} 

The great advantage of the concept of GPDs certainly is a unifying framework
for exclusive and inclusive reactions. The GPDs have a sound theoretical 
basis, their definition in QCD as Fourier transforms of hadronic matrix
elements is a direct generalization of the one for ordinary PDFs, their
evolution is known, the relationship to form factors, parton densities 
is well-established\cite{Muller:1994fv,Ji:1997ek,Radyushkin:1997ki}. For the
key reactions involving GPDs factorization theorems have been worked 
out\cite{Collins:1997fb,Radyushkin:1997ki}. Justifiably, these elements of the
formalism can already be considered {\it `future textbook knowledge'} on 
hard reactions.

A representation of GPDs as overlaps of light-cone wave functions allows a
clear interpretation\cite{Diehl:2001xz}. Whereas ordinary PDFs are parton 
densities, i.e.~probabilities, the GPDs
constitute interference amplitudes between different kinematical
situations. The information carried by GPDs thus exceeds the one of PDFs, just
as there is more to be learned from an amplitude then from its absolute 
square. Not only
longitudinal degrees of freedom, but also transverse d.o.f.~are accessible;
GPDs know about the orbital angular momentum of partons inside hadrons, which
is obvious from the hadron-helicity changing matrix elements and the
corresponding distribution functions, which are not present in the forward
case of ordinary PDFs. Moreover, the $q\bar q$ components of the nucleon can 
be directly studied from a kinematical domain again not accessible in forward
reactions (for more details see the contribution by {\it M.~Diehl}). 

\begin{figure}[htb]
\begin{center} 
\includegraphics[width=0.8\textwidth]{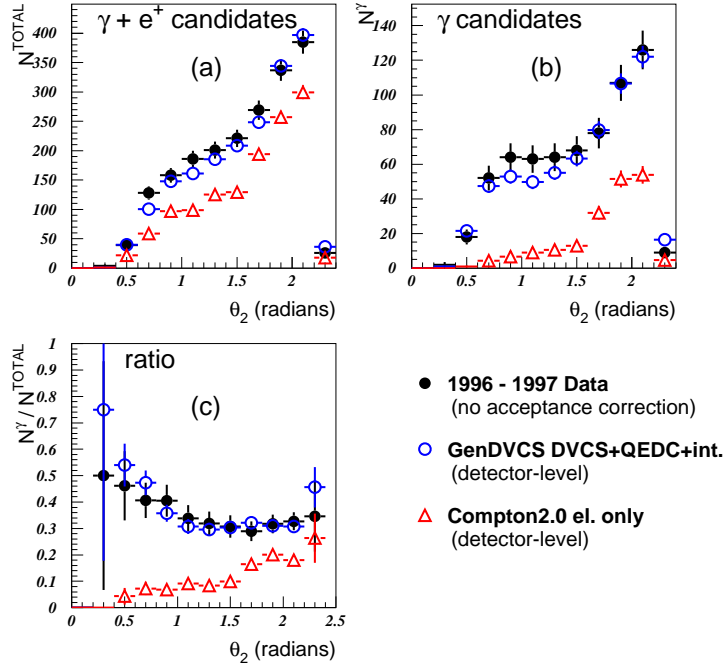}
\end{center} 
\caption{Observation of the DVCS reaction $e^+ p \rightarrow e^+ \gamma p$
at HERA in data taken with the ZEUS detector (see {\it L.~Favart}).
\label{fig:dvcs-ZEUS}}
\end{figure} 

\begin{figure}[htb]
\begin{center} 
\includegraphics[width=0.45\textwidth]{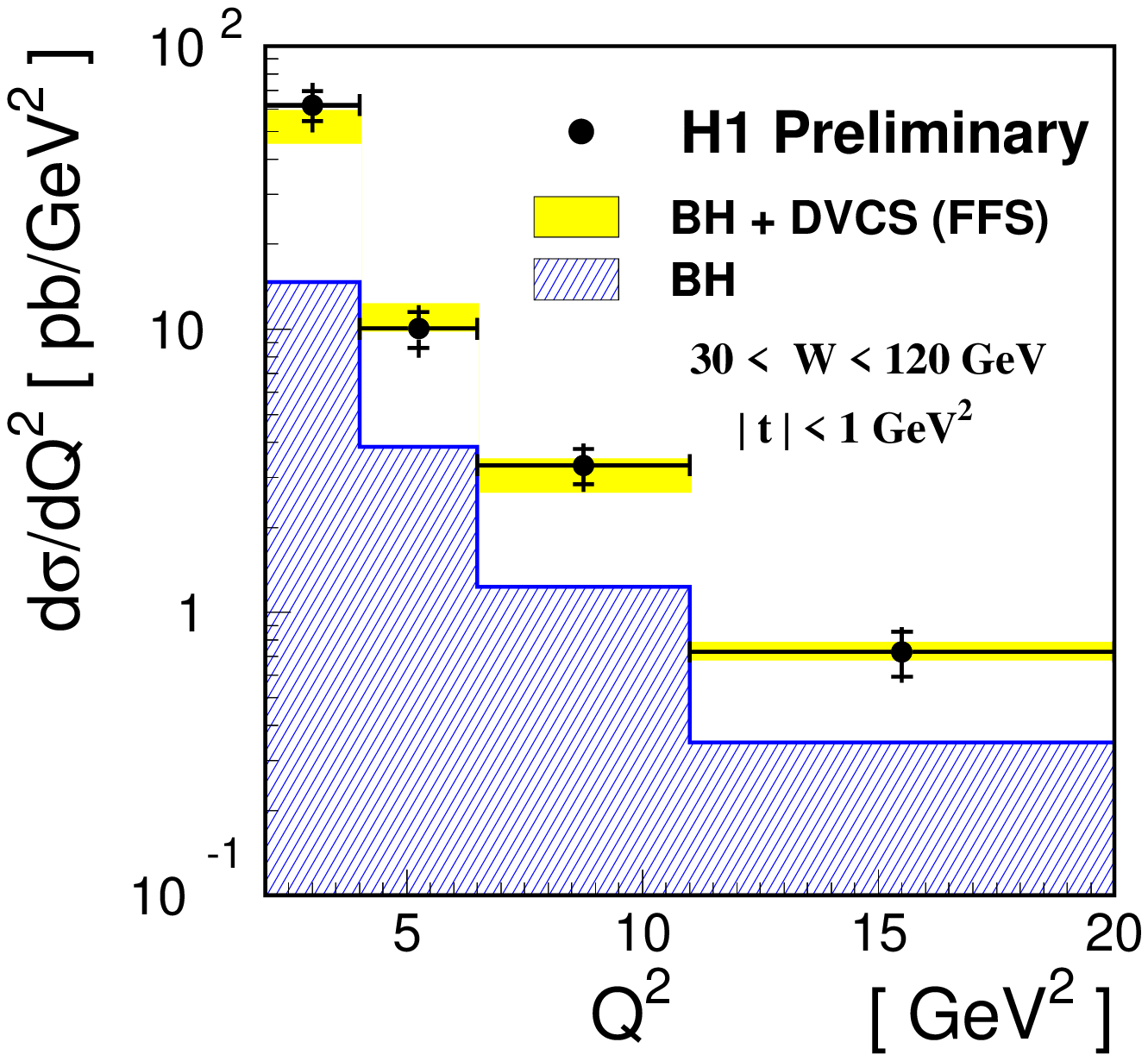}\quad
\includegraphics[width=0.45\textwidth]{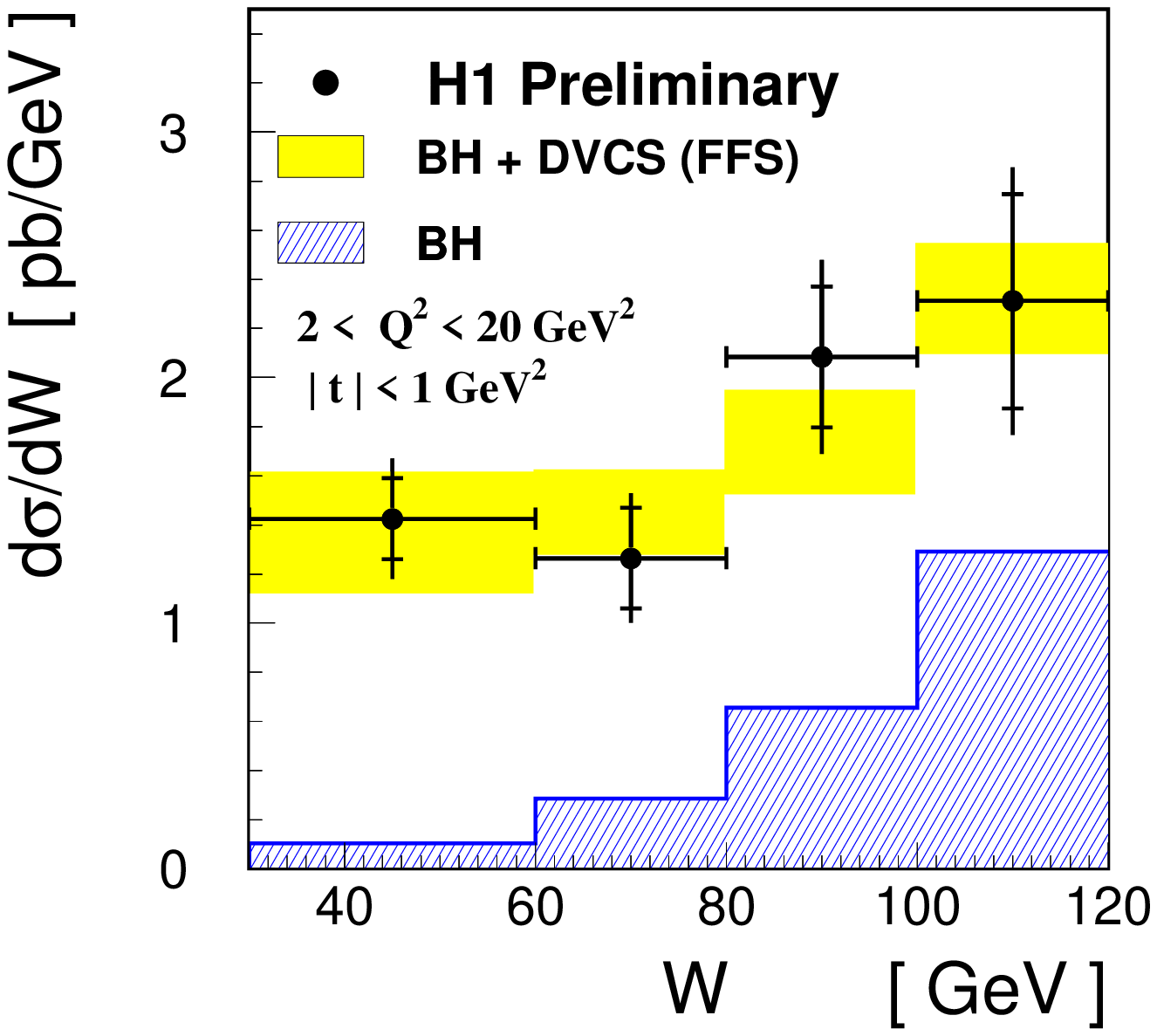}
\end{center} 
\caption{Differential cross section measurements for the DVCS reaction 
  $e^+ p \rightarrow e^+ \gamma p$ from H1 as a function of
  $Q^2$ (left) and $W$ (right). 
  The hatched histogram shows the contribution of the Bethe-Heitler 
  process to the reaction.
(see {\it L.~Favart})
\label{fig:dvcs-H1}}
\end{figure}

\subsection{Observability of hard exclusive reactions in the deeply 
virtual domain}

Surely, there is a price to pay for the rich information contained in
GPDs. These are objects depending on three variables (and have an additional
logarithmic scale dependence). Two variables are controlled by external
momenta, the momentum transfer squared $-t$, and the analogue of the Bjorken
variable $\xi$. The third variable $x$ constitutes an internal d.o.f.~of the
process and is integrated over, which presents an additional 
challenge, since observables will have to be deconvoluted to access the GPDs.  

In addition, hard exclusive reactions involving GPDs are difficult to measure,
since counting rates typically drop down drastically with increase of the
hardness of the process. Nevertheless, there is great progress on the
experimental side. Several observations have been reported on the key
processes to GPDs, like the DVCS signal from ZEUS and the DVCS cross section
measurement at H1 (see fig.\ \ref{fig:dvcs-ZEUS} and fig.\ \ref{fig:dvcs-H1}
and the contribution by {\it L.~Favart})

\begin{figure}[htb]
\begin{center}
\includegraphics[width=0.45\textwidth]{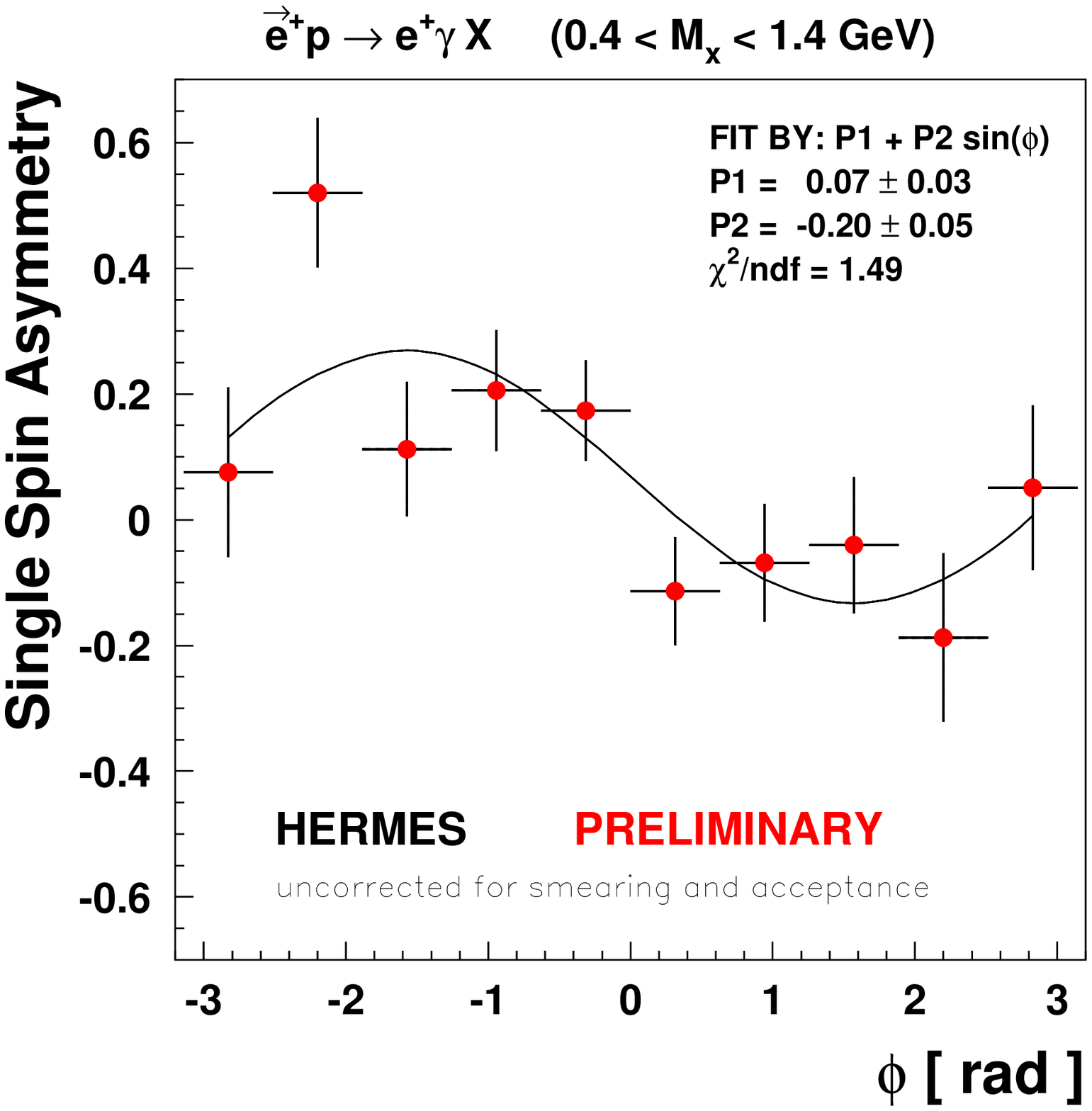}\quad
\includegraphics[width=0.45\textwidth]{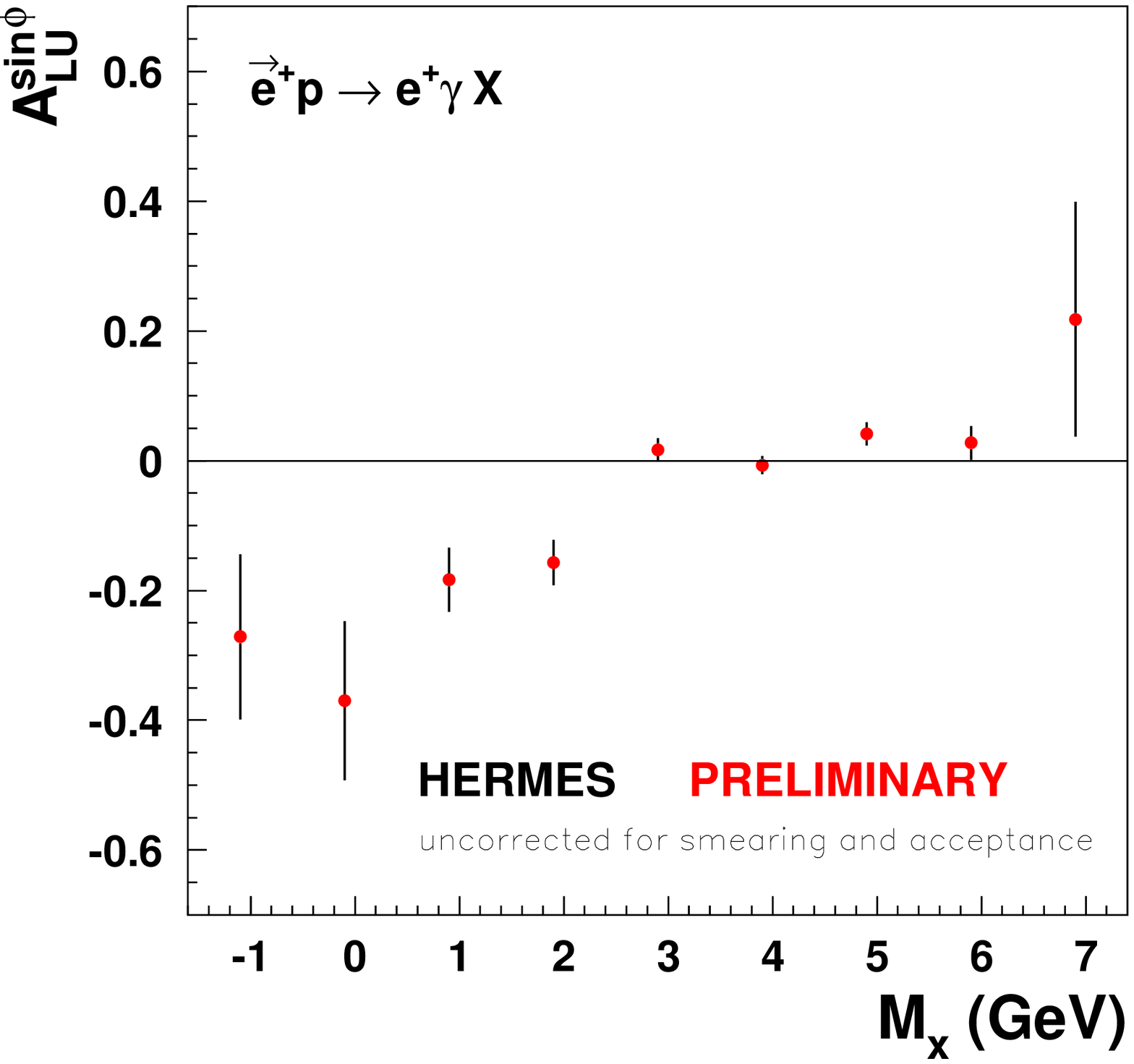}
\end{center}
\caption{The DVCS Single Spin Asymmetry from HERMES. Left: Beam-spin 
  asymmetry $A_{LU}$ for hard electroproduction of 
  photons as a function of the azimuthal angle $\phi$. The data correspond 
  to the missing mass region between 0.4 and 1.4 GeV. 
  Right: The beam-spin analyzing power $A_{LU}^{\sin}$ for hard 
  electroproduction of photons on hydrogen as a function of the 
  missing mass (see {\it J.~Volmer}).
\label{fig:dvcs-HERMES}}
\end{figure}

Even azimuthal asymmetries in the DVCS process have been already reported. In
fig.~\ref{fig:dvcs-HERMES} the beam-spin asymmetry observed by HERMES 
is displayed, together
with a $\sin(\phi)$-moment of the cross-section showing a clear non-vanishing
effect in dependence on the missing mass. Though the final state could not be
identified, the value of the asymmetry close to the proton mass can be 
assigned mainly to the Compton process. This result is very promising in view 
of improved particle identification available soon at 
HERMES (see {\it J.~Volmer}). 

\begin{figure}[ht] 
\begin{center} 
\includegraphics[width=0.45\textwidth]{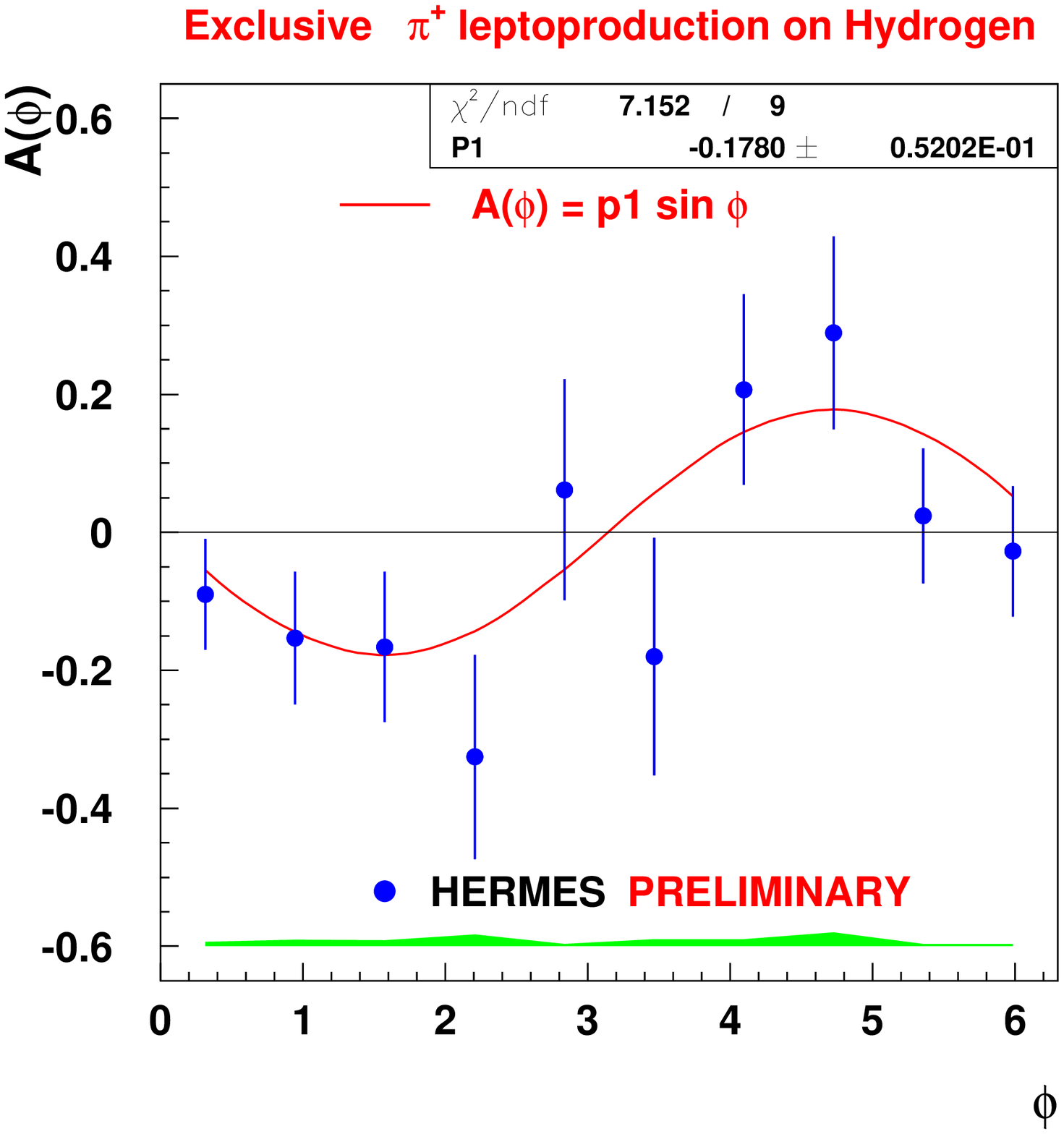}\quad
\includegraphics[width=0.45\textwidth]{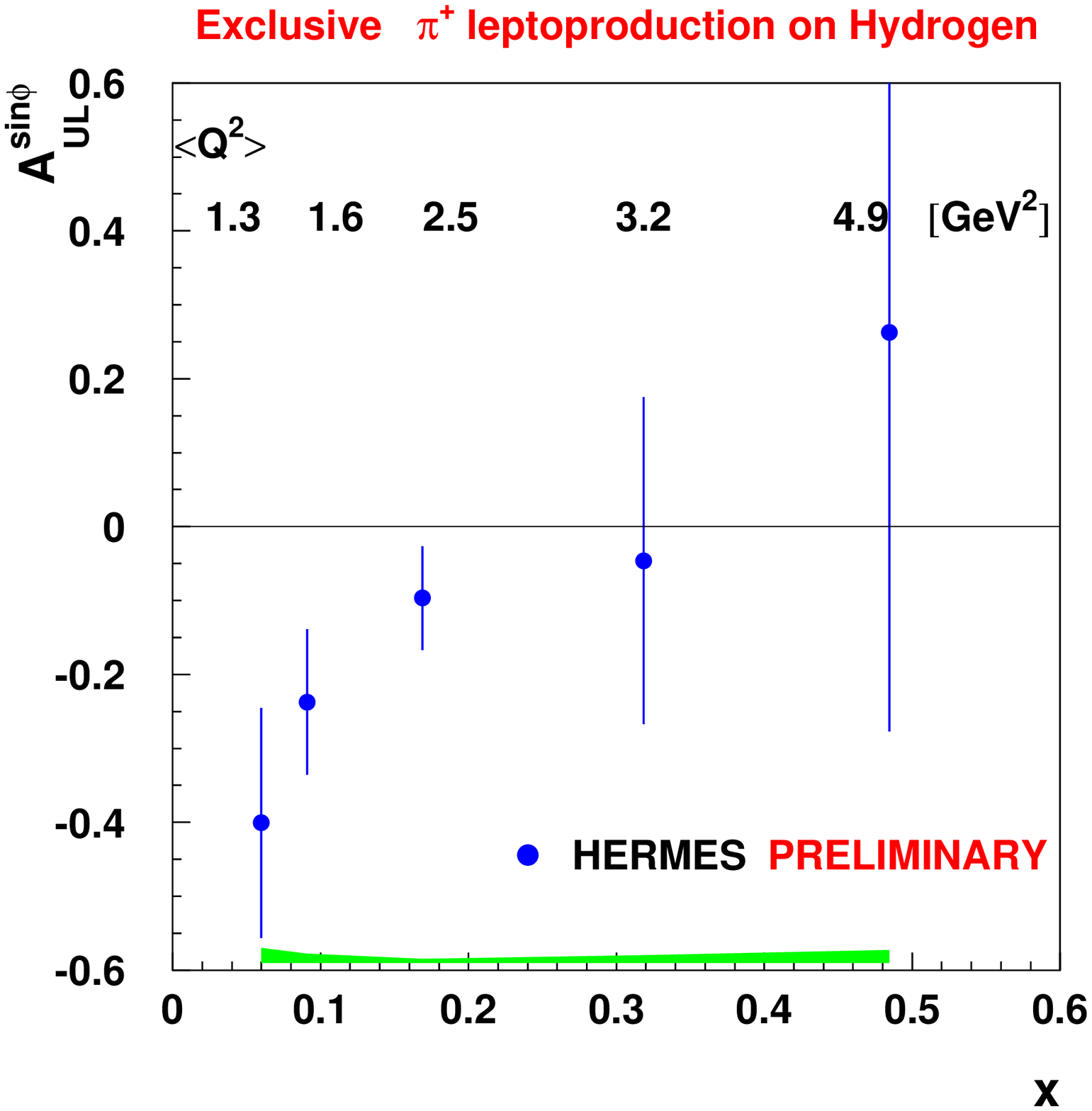}
\end{center} 
\caption{Azimuthal asymmetry in exclusive $\pi^+$ leptoproduction. Left: The
  asymmetry vs.~$\phi$. Right: The $\sin(\phi)$ moment of the 
polarized cross section vs.~$x$ (see {\it E.~Thomas}).
\label{fig:exclpi}}
\end{figure}

Similarly at HERMES an azimuthal asymmetry in hard exclusive leptoproduction
of $\pi^+$  has been observed. In fig.\ \ref{fig:exclpi} the asymmetry is 
shown together with the $\sin(\phi)$-moment vs.~$x$ (see {\it E.~Thomas}).

Only about three years ago, it was general believe that exclusive reactions in
the deeply virtual kinematical domain were forbiddingly difficult 
to measure. All the above mentioned findings prove the 
feasibility of those measurements. The more can be expected from dedicated
experiments or even dedicated accelerators like ELFE (see contribution 
by {\it D.~Ryckbosch}) or EIC (see {\it A.~Desphande}) tuned to the 
special needs of the observation of hard exclusive processes.

\subsection{Some aspects of the theoretical status of GPDs}

Likewise, on the theoretical side there has been great progress. It has been
shown that charge, spin and azimuthal asymmetries with longitudinal and
transversely polarized targets allow to disentangle all twist-2 and twist-3
GPDs when use is made of the interference with the Bethe-Heitler 
process\cite{Diehl:1997bu}
(see {\it D.~M\"uller}). NLO $\alpha_S$ corrections and evolution of GPDs 
from two-loop diagrams have been calculated\cite{Freund:2001rk}. From 
the results the 
interesting physical pictures arises that contrary to expectation 
meson-like excitations still contribute at small $x$-values, and at 
large $x$-values the sea-quark contributions are non-negligible, which
seems to indicate that even in the valence region one does not necessarily
strike the valence quarks directly (see {\it A.~Freund}).

Probably the strongest point in the theoretical concept of GPDs is the fact
that they link so many different processes. Key reactions are certainly the
DVCS and hard electroproduction of mesons. But also large momentum transfer
reactions are sensitive to GPDs, where they contribute through $\xi=0$
moments. The information obtainable from large momentum transfer exclusive
processes is complementary to the one from deeply virtual reactions.
Newly defined Compton form factors reflect the large $t$ dependence of the 
GPDs, similarly as elastic form factors do. With models for light-cone wave
functions one can achieve not only good agreement with elastic form factor and
Real Compton Scattering (RCS) data, but also arrive at interesting 
predictions for polarized and unpolarized RCS, VCS 
and electroproduction of mesons at large momentum 
transfer\cite{Diehl:1999kh} (see {\it P.~Kroll}).

Another option to access GPDs is the inverse DVCS, or timelike CS,
process $\gamma p \to \mu^+ \mu^- p$  with a large invariant mass of the 
lepton pair. It was shown that the process can be described in a 
factorized form similar to DVCS, and numerical estimates for its cross section
have been given (see {\it M.~Diehl}).

The most intriguing aspect in the formalism is the information carried by the
GPDs on orbital angular momentum. Actually, Ji's proposal\cite{Ji:1997ek} on 
a gauge invariant decomposition of the nucleon spin initiated the renewed
great interest in deeply virtual exclusive reactions. There is some 
theoretical debate on the issue: it appears that the elements in one possible
decomposition of the nucleon spin can be measured, but have no direct parton
representation, or in a different decomposition, the terms are separately 
interaction independent, gauge invariant, integrals over parton densities, but
there is no known way to measure them\cite{Jaffe:2001kr}. 

Anyway, there is general agreement that the information on orbital angular
momentum of partons is carried by GPDs; the situation is a challenge to
theory to find the appropriate way to isolate the desired information.
A contribution to the ongoing debate is the development of a physical picture
of the transverse localization of orbital angular momentum. Let $\Delta_T$ be
the transverse part of the momentum transfer in the process. The 
measurement of the $\Delta_T$ dependence of amplitudes can be inverted 
by Fourier transform to find the spatial impact parameter $b_T$ location 
of the partons. The transverse structure is directly observable when 
amplitudes are measured by interference (see {\it J.~Ralston}).

\subsection{GPD perspectives}

The exploration of the concept of Generalized Parton Distributions has just
begun. The theoretical sound basis has been carried to the NLO both 
in $\alpha_s$ corrections and evolution, and in higher-twist corrections.
First experimental observations on DVCS cross sections, asymmetries and
hard exclusive mesonproduction have confirmed the feasibility of measurements.
 
Spin physics will serve as the most important tool to access the 
different GPDs in a variety of asymmetries in many different reactions. 
The interplay between spin degrees of freedom and orbital 
angular momentum of partons in building up the nucleon spin will be the key 
to understand the inner structure of hadrons.

\section{Conclusion}

The physics goals in the field of hadron physics are clearly defined.
Experiments have already successfully provided a great deal of information
on the nucleon spin structure, and have proven to be capable to explore the
still missing elements and new theoretical concepts. We are looking forward to
an exciting future.

\section*{Acknowledgments}
This work has been supported by the TMR network
HPRN-CT-2000-00130.


\begin{thebibliography}{99}

\bibitem{Ashman:1988hv}
J.~Ashman {\it et al.}  [European Muon Collaboration],
Phys.\ Lett.\ B {\bf 206} (1988) 364.

\bibitem{Aubert:1974js}
J.~J.~Aubert {\it et al.},
Phys.\ Rev.\ Lett.\  {\bf 33} (1974) 1404;\newline
J.~E.~Augustin {\it et al.},
Phys.\ Rev.\ Lett.\  {\bf 33} (1974) 1406.

\bibitem{Fukuda:1998mi}
Y.~Fukuda {\it et al.}  [Super-Kamiokande Collaboration],
Phys.\ Rev.\ Lett.\  {\bf 81} (1998) 1562.

\bibitem{SPI} SPIRES TOPCITE 1000+ Database.

\bibitem{Filippone:2001ux}
For a recent review see 
B.~W.~Filippone and X.~Ji,
hep-ph/0101224.

\bibitem{Ralston:1979ys}
J.~P.~Ralston and D.~E.~Soper,
Nucl.\ Phys.\ B {\bf 152} (1979) 109;\newline
%
R.~L.~Jaffe,
hep-ph/9602236.

\bibitem{Barone:2001sp}
For a recent review see
V.~Barone, A.~Drago and P.~G.~Ratcliffe,\newline
hep-ph/0104283.

\bibitem{Airapetian:2000tv}
A.~Airapetian {\it et al.}  [HERMES Collaboration],
Phys.\ Rev.\ Lett.\  {\bf 84} (2000) 4047;
%
A.~Airapetian {\it et al.}  [HERMES Collaboration],
hep-ex/0104005;
%
A.~Bravar  [Spin Muon Collaboration],
Nucl.\ Phys.\ Proc.\ Suppl.\  {\bf 79} (1999) 520.

\bibitem{Goeke:2001tz}
For a recent review see
K.~Goeke, M.~V.~Polyakov and M.~Vanderhaeghen,
hep-ph/0106012.

\bibitem{Airapetian:2001yk}
A.~Airapetian {\it et al.} [HERMES Collaboration],
hep-ex/0106068;\newline
%
C.~Adloff {\it et al.}  [H1 Collaboration],
hep-ex/0107005;\newline
%
P.~R.~Saull {\it et al.} [ZEUS Collaboration],
hep-ex/0003030;\newline
%
S.~Stepanyan {\it et al.}  [CLAS Collaboration],
hep-ex/0107043.

\bibitem{Bianchi:1999qs}
N.~Bianchi and E.~Thomas,
Phys.\ Lett.\ B {\bf 450} (1999) 439.

\bibitem{Abe:1996dc}
K.~Abe {\it et al.}  [E143 Collaboration],
Phys.\ Rev.\ Lett.\  {\bf 76} (1996) 587;\newline
%
P.~L.~Anthony {\it et al.}  [E155 Collaboration],
Phys.\ Lett.\ B {\bf 458} (1999) 529.

\bibitem{Leader:1998qv}
E.~Leader, A.~V.~Sidorov and D.~B.~Stamenov,
Phys.\ Rev.\ D {\bf 58} (1998) 114028.

\bibitem{deFlorian:2000bm}
D.~de Florian and R.~Sassot,
Phys.\ Rev.\ D {\bf 62} (2000) 094025.

\bibitem{Adeva:1998qz}
B.~Adeva {\it et al.}  [Spin Muon Collaboration],
Phys.\ Lett.\ B {\bf 420} (1998) 180;
%
K.~Ackerstaff {\it et al.}  [HERMES Collaboration],
Phys.\ Lett.\ B {\bf 464} (1999) 123.

\bibitem{Airapetian:2000ib}
A.~Airapetian {\it et al.}  [HERMES Collaboration],
Phys.\ Rev.\ Lett.\  {\bf 84} (2000) 2584.

\bibitem{Jaffe:1997yz}
R.~L.~Jaffe,
hep-ph/9710465.

\bibitem{Collins:1993kk}
J.~Collins,
Nucl.\ Phys.\ B {\bf 396} (1993) 161.

\bibitem{Mulders:1996dh}
P.~J.~Mulders and R.~D.~Tangerman,
Nucl.\ Phys.\ B {\bf 461} (1996) 197
[Erratum-ibid.\ B {\bf 484} (1996) 538];\newline
%
D.~Boer, R.~Jakob and P.~J.~Mulders,
Nucl.\ Phys.\ B {\bf 564} (2000) 471.

\bibitem{Boer:2001he}
D.~Boer,
Nucl.\ Phys.\ B {\bf 603} (2001) 195.

\bibitem{Henneman:2001ev}
A.~A.~Henneman, D.~Boer and P.~J.~Mulders,
hep-ph/0104271.

\bibitem{Anselmino:2001vs}
M.~Anselmino, D.~Boer, U.~D'Alesio and F.~Murgia,
Phys.\ Rev.\ D {\bf 63} (2001) 054029.

\bibitem{Bacchetta:2000jk}
A.~Bacchetta and P.~J.~Mulders,
Phys.\ Rev.\ D {\bf 62} (2000) 114004.

\bibitem{Collins:1994kq}
J.~C.~Collins, S.~F.~Heppelmann and G.~A.~Ladinsky,
Nucl.\ Phys.\ B {\bf 420} (1994) 565;
%
J.~C.~Collins and G.~A.~Ladinsky,
hep-ph/9411444;
%
X.~Artru and J.~Collins,
Z.\ Phys.\ C {\bf 69} (1996) 277;
%
R.~L.~Jaffe, X.~Jin and J.~Tang,
Phys.\ Rev.\ Lett.\  {\bf 80} (1998) 1166;
%
A.~Bianconi, S.~Boffi, R.~Jakob and M.~Radici,
Phys.\ Rev.\ D {\bf 62} (2000) 034008.


\bibitem{Kanazawa:2001cx}
Y.~Kanazawa and Y.~Koike,
Phys.\ Rev.\ D {\bf 64} (2001) 034019.

\bibitem{Muller:1994fv}
D.~Muller, D.~Robaschik, B.~Geyer, F.~M.~Dittes and J.~Horejsi,
Fortsch.\ Phys.\  {\bf 42} (1994) 101.

\bibitem{Ji:1997ek}
X.~Ji,
Phys.\ Rev.\ Lett.\  {\bf 78} (1997) 610.

\bibitem{Radyushkin:1997ki}
A.~V.~Radyushkin,
Phys.\ Rev.\ D {\bf 56} (1997) 5524.

\bibitem{Collins:1997fb}
J.~C.~Collins, L.~Frankfurt and M.~Strikman,
Phys.\ Rev.\ D {\bf 56} (1997) 2982;
%
X.~Ji and J.~Osborne,
Phys.\ Rev.\ D {\bf 58} (1998) 094018;
%
J.~C.~Collins and A.~Freund,
Phys.\ Rev.\ D {\bf 59} (1999) 074009.

\bibitem{Diehl:2001xz}
M.~Diehl, T.~Feldmann, R.~Jakob and P.~Kroll,
Nucl.\ Phys.\ B {\bf 596} (2001) 33
[Erratum-ibid.\ B {\bf 605} (2001) 647].

\bibitem{Diehl:1997bu}
M.~Diehl, T.~Gousset, B.~Pire and J.~P.~Ralston,
Phys.\ Lett.\ B {\bf 411} (1997) 193;
%
A.~V.~Belitsky, D.~Muller, L.~Niedermeier and A.~Schafer,
Nucl.\ Phys.\ B {\bf 593} (2001) 289;
%
A.~V.~Belitsky and D.~Muller,
Nucl.\ Phys.\ B {\bf 589} (2000) 611;
%
A.~V.~Belitsky, A.~Kirchner, D.~Muller and A.~Schafer,
Phys.\ Lett.\ B {\bf 510} (2001) 117.

\bibitem{Freund:2001rk}
A.~Freund and M.~F.~McDermott,
hep-ph/0106319.

\bibitem{Diehl:1999kh}
M.~Diehl, T.~Feldmann, R.~Jakob and P.~Kroll,
Eur.\ Phys.\ J.\ C {\bf 8} (1999) 409;
%
M.~Diehl, T.~Feldmann, R.~Jakob and P.~Kroll,
Phys.\ Lett.\ B {\bf 460} (1999) 204;
%
H.~W.~Huang and P.~Kroll,
Eur.\ Phys.\ J.\ C {\bf 17} (2000) 423.


\bibitem{Jaffe:2001kr}
R.~L.~Jaffe,
Phil.\ Trans.\ Roy.\ Soc.\ Lond.\ A {\bf 359} (2001) 391.

\end{thebibliography}
\end{document}